\documentclass[aps,prl,twocolumn,superscriptaddress, runinaddress, showpacs]{revtex4}

\usepackage{graphicx}

\begin{document}


\title{Bulk and surface-sensitive high-resolution photoemission study \\of Mott-Hubbard systems SrVO$_3$ and CaVO$_3$}




\author{R.~Eguchi}
\affiliation{Institute for Solid State Physics (ISSP), University of Tokyo, Kashiwanoha, Kashiwa, Chiba 277-8581, Japan}
\affiliation{RIKEN SPring-8 Center, Sayo-cho, Sayo-gun, Hyogo 679-5148, Japan}

\author{T.~Kiss}
\affiliation{Institute for Solid State Physics (ISSP), University of Tokyo, Kashiwanoha, Kashiwa, Chiba 277-8581, Japan}

\author{S.~Tsuda}
\affiliation{Institute for Solid State Physics (ISSP), University of Tokyo, Kashiwanoha, Kashiwa, Chiba 277-8581, Japan}

\author{ T.~Shimojima}
\affiliation{Institute for Solid State Physics (ISSP), University of Tokyo, Kashiwanoha, Kashiwa, Chiba 277-8581, Japan}

\author{ T.~Mizokami}
\affiliation{Institute for Solid State Physics (ISSP), University of Tokyo, Kashiwanoha, Kashiwa, Chiba 277-8581, Japan}

\author{T.~Yokoya}
\affiliation{Institute for Solid State Physics (ISSP), University of Tokyo, Kashiwanoha, Kashiwa, Chiba 277-8581, Japan}

\author{A.~Chainani}
\affiliation{RIKEN SPring-8 Center, Sayo-cho, Sayo-gun, Hyogo 679-5148, Japan}

\author{S.~Shin}
\affiliation{Institute for Solid State Physics (ISSP), University of Tokyo, Kashiwanoha, Kashiwa, Chiba 277-8581, Japan}
\affiliation{RIKEN SPring-8 Center, Sayo-cho, Sayo-gun, Hyogo 679-5148, Japan}

\author{I. H.~Inoue}
\affiliation{Correlated Electron Reserch Center (CERC), National Institute of Advance Industrial Science and Tecnology (AIST), Tsukuba Central 4, Tsukuba 305-8562, Japan}

\author{T.~Togashi}
\affiliation{RIKEN SPring-8 Center, Sayo-cho, Sayo-gun, Hyogo 679-5148, Japan}

\author{S.~Watanabe}
\affiliation{Institute for Solid State Physics (ISSP), University of Tokyo, Kashiwanoha, Kashiwa, Chiba 277-8581, Japan}

\author{C. Q.~Zhang}
\affiliation{Beijing Center for Crystal R\&D, Chinese Academy of Science, Zhongguancun, Beijing 100080, China}

\author{C. T.~Chen}
\affiliation{Beijing Center for Crystal R\&D, Chinese Academy of Science, Zhongguancun, Beijing 100080, China}

\author{M.~Arita}
\affiliation{Hiroshima Synchrotron Radiation Center, Hiroshima University, Higashi-Hiroshima 739-8526, Japan}

\author{K.~Shimada}
\affiliation{Hiroshima Synchrotron Radiation Center, Hiroshima University, Higashi-Hiroshima 739-8526, Japan}

\author{H.~Namatame}
\affiliation{Hiroshima Synchrotron Radiation Center, Hiroshima University, Higashi-Hiroshima 739-8526, Japan}

\author{M.~Taniguchi}
\affiliation{Hiroshima Synchrotron Radiation Center, Hiroshima University, Higashi-Hiroshima 739-8526, Japan}
\affiliation{Graduate School of Science, Hiroshima University, Higashi-Hiroshima 739-8526, Japan}


\date{\today}

\begin{abstract}
We study the electronic structure of Mott-Hubbard systems SrVO$_{3}$ and CaVO$_3$ with bulk and surface-sensitive high-resolution photoemission spectroscopy (PES), using a VUV laser, synchrotron radiation and a discharge lamp ($h\nu$ = 7 - 21 eV).  A systematic suppression of the density of states (DOS) within $\sim$ 0.2 eV of the Fermi level ($E_F$) is found on decreasing photon energy i.e. on increasing bulk sensitivity. The coherent band in SrVO$_{3}$ and CaVO$_3$ is shown to consist of surface and bulk derived features, separated in energy.  The stronger distortion on surface of CaVO$_{3}$ compared to SrVO$_{3}$ leads to higher surface metallicity in the coherent DOS at $E_F$, consistent with recent theory. 

\end{abstract}

\pacs{71.27.+a, 71.20.Be, 79.60.-i}

\maketitle

The electronic properties of 3$d$ transition metal oxides with strong electron correlations have been extensively studied, particularly since the discovery of high $Tc$ superconductivity and colossal magnetoresistance \cite{dagotto, imada}. Carrier doping and/or bandwidth control leads to attractive and complicated phase transitions, such as superconducting and metal-insulator transitions (MIT) in various perovskite-type compounds \cite{dagotto, imada}. Such MIT are known to occur systematically by filling control due to the closing of a Mott-Hubbard (MH) type gap. According to MH theory \cite{Hubbard}, such a MIT is controlled by the relative magnitude of on-site Coulomb interaction $U$ and bandwidth $W$. Valence band photoemission spectroscopy (PES) studies validate this picture \cite{fujimori}. The dynamical mean-field theory (DMFT) has helped to elucidate the ultraviolet photoemission spectroscopy (UPS) and optical conductivity of the MH MIT \cite{zhang, roz}.  It is now well-known that strongly correlated materials exhibit a 
coherent and incoherent band and presently, DMFT is considered the most powerful theoretical method to elucidate their electronic structure (ES).

The perovskite-type 3$d^1$ configuration metals Ca$_{1-x}$Sr$_x$VO$_3$ 
are typical materials whose valence band spectra are compared with DMFT. This is because
 it is believed that $d$-bandwidth is controlled by the V-O-V bond angle, which changes from 180$^\circ$ for cubic SrVO$_3$ to 160$^\circ$ for orthorhombic CaVO$_3$ \cite{cham} i.e., the $d$-bandwidth $W$ of SrVO$_3$ is larger than that of CaVO$_3$ \cite{inoueB}, while on-site $U$ is similar for $d^1$ (V$^{4+}$) oxides. Early PES and inverse-PES of Ca$_{1-x}$Sr$_x$VO$_3$ in the V 3$d$ region showed a three-peak structure \cite{morikawa95}, namely, the coherent band (quasi-particle band) at $E_F$ and the incoherent band corresponding to upper and lower Hubbard bands, located at a few eV above and below $E_F$. UPS experiments were also interpreted to show systematic spectral weight transfer from coherent band to the incoherent band with increasing $U/W$, consistent with early DMFT studies \cite{inoueL}. However, a very recent LDA+DMFT study shows that SrVO$_{3}$ and CaVO$_3$ are very similar, consistent with thermodynamic and bulk-sensitive experiments \cite{Nekrasov}. In particular, it was clarified that the V-O-V bond angle change involves (i) a decreasing $d$-$p$-$d$ hybridization and (ii) a compensating, increasing $d$-$d$ hybridization, resulting in  a very small change in $t_{2g}$ bandwidth \cite{Nekrasov}.

The bulk and surface contributions in the ES of Ca$_{1-x}$Sr$_x$VO$_3$ have been studied using photon energy dependent PES \cite{maiti2}. The surface sensitivity of PES can be varied by changing the exciting photon energy, since the mean free path (MFP) of electrons depends sensitively on their kinetic energy \cite{seah}. 
Based on a $U/W$ analysis of spectral weight transfer between the coherent and incoherent features, it was concluded that the bulk and surface ES of SrVO$_3$ is metallic, while the surface of CaVO$_3$ is a MH insulator and its bulk is metallic \cite{maiti2}. This analysis does not take into account the possibility of separate features due to surface and bulk ES within the coherent band.  DMFT studies show that the reduced coordination number leads to a charge transfer of 0.06 electrons from the surface $d_{xy}$ band to the surface $d_{xz}$, $d_{yz}$ bands \cite{lie}. The effective mass gets enhanced at the surface with a narrowing of the coherent band, effectively leading to higher DOS at $E_F$. 
Further, recent bulk-sensitive PES measurements of CaVO$_{3}$ and SrVO$_3$ using soft X-ray (SX) synchrotron radiation ($h\nu$ = 900 eV) \cite{sekiyama} have shown that V $3d$ DOS at $E_F$ is similar for both systems, indicating absence of spectral weight transfer between coherent and incoherent features, as explained in detail in \cite{Nekrasov}. In very profound studies of $d^{1}$ MH titanate systems \cite{ohtomo},  recent experiments and theory have established `electronic reconstruction' leading to a metallic surface of a MH insulator, and a metallic interface between a MH insulator and band insulator. This behavior is also derived from charge density transferred to the surface/interface \cite{ohtomo}.

Thus, there exist discrepancies of the results in terms of surface versus bulk
ES of CaVO$_{3}$ and SrVO$_3$ that need to be addressed. 
In this work, we investigate surface versus bulk ES within the coherent band of CaVO$_{3}$ and SrVO$_3$. We adopt a different route to study surface and bulk ES : low energy PES, simply because the universal MFP curve has a minimum at about 30 - 50 eV electron kinetic energy \cite{seah}. Along with higher bulk sensitivity, we get the added advantage of very high resolution, using a new laser system ($h\nu$ = 6.994 eV), low-energy ($h\nu$ = 9-13 eV) synchrotron radiation and a discharge lamp ($h\nu$ = 21.218 eV). The photoelectron MFP at $h\nu$ = 6.994 eV, the lowest photon energy used in the present study, is $\sim$100 $\AA$ \cite{seah}. This is much more bulk sensitive than previous PES measurements done with Al $K\alpha$ ($h\nu$ = 1486.6 eV, MFP $\approx$ 20 $\AA$) or soft x-rays ($h\nu$ = 900 eV, MFP $\approx$ 15 $\AA$). At the highest energy used here ($h\nu$ = 21.218 eV), the MFP is $\approx$ 5 $\AA$, and hence we progressively span from very surface-sensitive to very bulk-sensitive measurements.

PES experiments were carried out using (i) a Scienta SES2002 PES spectrometer and a GAMMADATA He discharging lamp with a monochromator, (ii) a wide-angle-lens ($\pm$ 14$^o$) Scienta R4000 PES spectrometer and a VUV laser at ISSP \cite{kiss}. The VUV laser system, developed recently \cite{togashi}, employs the 6th harmonic light of energy 6.994 eV, obtained by using the second harmonic KBe$_2$BO$_3$F$_2$ crystal with a frequency tripled single-mode Nd:YVO$_4$ laser. The intensity is more than 10$^{15}$ photons per second with a line width of 0.25 meV. 
PES experiments were also carried out using Scienta SES2002 PES spectrometer and synchrotron radiation 
($h\nu$ = 13.8 eV and 9.2 eV) at BL-9A, at HiSOR \cite{tanimatsuari}. All systems use a flowing liquid He cryostat with a thermally shielded sample holder. Single crystals of SrVO$_3$ and CaVO$_3$ samples were fractured $in$-$situ$ to obtain a clean surface. PES measurements were carried out at a vacuum of $\sim$ 10$^{-10}$ Torr at 6 - 200 K. Since the photon energies are low, we ensured that all the spectra reported here are angle-integrated
by actually checking for angle-resolved effects by varying the incidence angle so as to cover the first Brillouin zone.
The energy resolutions were set to 8 meV to get reasonable signal intensity. The $E_F$ was referenced to that of a gold film evaporated onto the sample holder.

\begin{figure}
\includegraphics[width=0.4\textwidth]{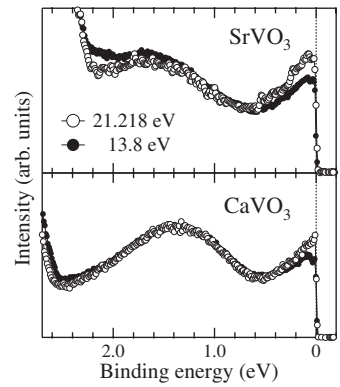}
\caption{\label{} Photoemission spectra of SrVO$_3$ (top) and CaVO$_3$ (bottom) measured using 21.218 eV and 13.8 eV photon energy at 6 K.}
\end{figure}

Figure 1 shows the comparison of spectra measured using 21.218 eV and 13.8 eV photon energy at 6 K 
for SrVO$_3$ and CaVO$_3$. The spectra are normalized for area under the curve. 
The 21.218 eV spectra are consistent with early work \cite{inoueL}. Small differences are seen in the lower binding energy coherent part, at and near $E_F$, which we investigate in detail.
Figure 2(a) and 2(b) show bulk sensitive PE spectra of SrVO$_3$ and CaVO$_3$ measured using laser source ($h\nu$ = 6.994 eV) at 6 K. The experimental spectra are normalized to match at the dip feature around 0.6 eV, assuming a similar behavior as in Fig. 1. This normalization suggests that spectral weight in the coherent feature is significantly different ($\sim$ 50 \% at $E_F$) between SrVO$_3$ and CaVO$_3$. However, since it is impossible to investigate the lower Hubbard band with $h\nu$ = 6.994 eV, and the normalization used above does not separate out bulk versus surface contributions and an ambiguous background, it leads to a difference between SrVO$_3$ and CaVO$_3$. This is inconsistent with thermodynamic and bulk-sensitive experiments \cite{inoueB,Nekrasov,maiti2,sekiyama}.

More significantly, we observe that the peak of the coherent feature is not at $E_F$ [Fig. 2(b)], but at an energy of $\sim$ 0.2 eV in both SrVO$_3$ and CaVO$_3$. The intensity suppression observed at $E_F$ is more in laser excited spectra than in the spectra measured using $h\nu$ = 21.218 eV and 13.8 eV (Fig. 1). 
It suggests that this intensity suppression is more in the bulk, while the intensity at $E_F$ is higher in
the surface ES. 
This behavior within the coherent band has not been addressed in earlier studies. 
The ratio of the spectral intensity at $E_F$ to the coherent peak (at 0.18 eV binding energy) intensity is about 0.7 and 0.5 in SrVO$_3$ and CaVO$_3$, respectively. We have also checked the temperature dependence of this behavior.
Figures 2(c) and (d) show the temperature dependent spectra near $E_F$ measured using laser source at 6 K, 100 K, and 200 K. The spectral intensity suppression does not change with temperature.

\begin{figure}
\includegraphics[width=0.45\textwidth]{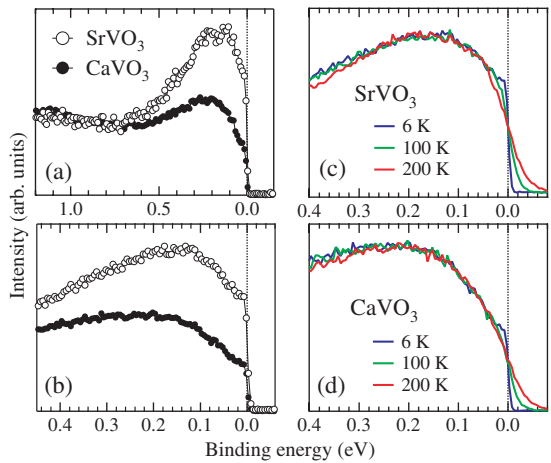}
\caption{\label{}Photoemission spectra of SrVO$_3$ and CaVO$_3$ measured using laser source ($h\nu$ = 6.994 eV) at 6 K. (b) Expanded view of the coherent band near $E_F$. (c) Temperature-dependent photoemission spectra of SrVO$_3$ and (d) CaVO$_3$ near $E_F$ at 6 K, 100 K, and 200 K.}
\end{figure}

In order to confirm and establish spectral changes as a function of photon energy, and hence bulk sensitivity, 
we compare high signal to noise ratio photon-energy-dependent spectra of SrVO$_3$ and CaVO$_3$ at and near $E_F$ measured using 21.218, 13.8, 9.2, and 6.994 eV photons in Fig. 3. The spectra for SrVO$_3$ and CaVO$_3$ are independently normalized at 0.5 eV binding energy so as to see the relative change with photon energy. The intensity at and near $E_F$ is systematically suppressed with lower photon energy. The relative spectral change is much larger in CaVO$_3$ than in SrVO$_3$. A clear structure is seen within 40 meV of $E_F$ in the surface sensitive spectra obtained with 21.218 and 13.8 eV, particularly for CaVO$_3$.

In order to estimate surface and bulk contributions within the coherent band at and near $E_F$, and approximate MFPs, we have carried out the following analysis : The normalized PE spectral intensity $I$($E$) is expressed as $I$($E$) = exp(-$d$/$\lambda$)$I_{b}$($E$)+[1-exp(-$d$/$\lambda$)]$I_{s}$($E$), where $I_{b}$ and $I_{s}$ is the normalized bulk and surface spectra, $d$ is the thickness of the surface layer, and $\lambda$ is the MFP, respectively \cite{liu}. We obtain $I_{b}$ and $I_{s}$ spectra, using the 21.218 eV and 6.994 eV excited spectra, and assuming $d$ = 7.5 $\AA$, $\lambda_{21.218}$ = 5 $\pm$ 1 $\AA$ and $\lambda_{6.994}$ = 85 $\pm$ 10 $\AA$.  The extracted  $I_{b}$ and $I_{s}$ spectra of SrVO$_3$ and CaVO$_3$ [see Fig. 3(a) and 3(b)] also consistently reproduce the spectra obtained with
$h\nu$ = 13.8 eV($\lambda_{13.8}$ = 11 $\pm$ 1 $\AA$) and $h\nu$ = 9.2 eV  ($\lambda_{9.2}$ = 23 $\pm$ 2 $\AA$). The results are in very good agreement with experimental spectra, including the fine structures below 0.1 eV-binding energy. See Ref. \cite{Eguchi} for a complete analysis. The error bars in estimating MFPs are approximately $\pm$10 - 20$\%$, but the qualitative trend is consistent with He I data being more surface sensitive and the lower photon energy($\sim$ 14 to 7 eV) data is progressively more bulk sensitive, consistent with the universal MFP curve \cite{seah}. The surface versus bulk analysis is done independently for SrVO$_3$ and CaVO$_3$, but for a relative comparison between them which eliminates photon energy dependence of cross-section, we normalize the intensities at the peak ($\equiv$ 0.18 eV) in the bulk-derived spectra in Figs. 3(a) and 3(b) (red horizontal line), assuming the analysis of ref. \onlinecite{sekiyama} as correct. This leads to a difference in the DOS at $E_F$ in the extracted bulk contribution of  $\sim$ 25 \% between SrVO$_3$ and CaVO$_3$, while Ref. \cite{sekiyama} estimates it as $\sim$ 15 \%. These values are consistent with thermodynamic \cite{inoueB} and bulk-sensitive experiments \cite{Nekrasov,maiti2,sekiyama}, as well as calculated quasiparticle weight $Z$ (or effective mass $m$*/$m$ = 1/$Z$, which varies between 2.1 - 2.4 \cite{Nekrasov, sekiyama} or 2.2 - 3.5 \cite{pavarini}. Thus, the normalization presented in Fig. 3 is appropriate for the bulk ES. On separating out the bulk component, the surface-sensitive spectra at $E_F$ exhibit higher metallicity in CaVO$_3$, and also in SrVO$_3$ but to a lesser extent, consistent with recent studies \cite{lie, ohtomo,nolt1}. In contrast with Ref. \cite{maiti2}, the surface ES of CaVO$_3$ is not a MH insulator.

\begin{figure}
\includegraphics[width=0.4\textwidth]{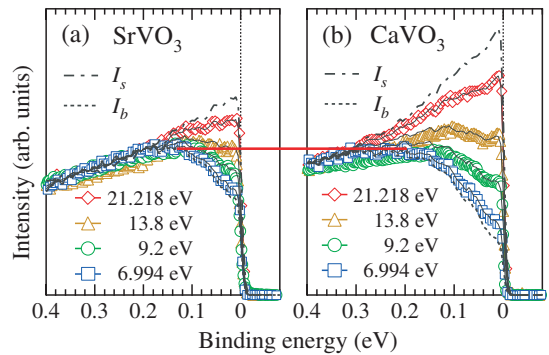}
\caption{\label{} Photon-energy-dependent photoemission spectra of (a) SrVO$_3$ and (b) CaVO$_3$ near $E_F$ measured using 21.218, 13.8, 9.2, and 6.994 eV photons at 6 K. The extracted the bulk and surface spectral function of SrVO$_3$ and CaVO$_3$ were also plotted (dotted and dotted-dashed lines) with the reproduced spectra (thin lines) of each excited energy.}
\end{figure}

The operative interactions responsible for the observations need careful discussion. First, the observed suppression in bulk-sensitive PES cannot be elucidated by band structure calculations, which indicate a maximum at $E_F$ in the occupied DOS \cite{take}. 
Secondly, while it is known that magnetic or charge fluctuations can lead to a pseudogap in the DOS \cite{pseudo}, a magnetic or CDW transition has not been observed in any experimental study of SrVO$_3$ and CaVO$_3$. The phonon DOS in SrVO$_3$ and CaVO$_3$ has also not been reported. In related materials like SrTiO$_3$, BaTiO$_3$, LaTiO$_3$, and YTiO$_3$, the optical phonon mode is observed at 67, 62, 66, and 70 meV, respectively \cite{katsucranbarker}. Thus, the 0.2 eV energy pseudogap-like feature in the present experiment is considered too large to be related to phonons. Moreover, there is no report indicating strong electron-phonon coupling in SrVO$_3$ and CaVO$_3$.

DMFT calculations for the bulk ES of SrVO$_3$ and CaVO$_3$ also predict a peak at $E_F$. 
However, DMFT calculations of the surface ES revealed strong modifications of the coherent and incoherent bands of $d^{1}$ MH systems \cite{lie, nolt1, ohtomo}. At the surface, the $t_{2g}$ degeneracy is lifted, leading to a separation of $d_{xy}$ and $d_{yz,zx}$ derived states. The $d_{xy}$ band lies in the plane and exhibits strong dispersion within the surface plane. The DOS of the more strongly correlated $d_{yz,zx}$ states is narrower, leading to an enhancement of its spectral weight at $E_F$. These changes are caused by electronic charge density being transferred to the surface. Experiments and theory show very similar results for the surface of a MH insulator and the interface ES of a MH and band insulator, which actually makes the surface/interface metallic \cite{ohtomo}.  We believe the feature within 40 meV of $E_F$ is the surface $d_{yz,zx}$ states, followed by the surface $d_{xy}$ states between 40 meV to $\sim$ 0.2 eV (Fig. 3). These are suppressed in the bulk, revealing the bulk $t_{2g}$ feature at $\sim$ 0.2 eV with a pseudogap at $E_F$.
The present experimental results demonstrate that the $t_{2g}$ degeneracy is indeed lifted on the surface, observed by increasing photon energy from 7 to 21 eV or increasing surface sensitivity. The relative change in CaVO$_3$ is significantly larger than in SrVO$_3$, and we attribute it to the difference between their surface ES. 
Recent DMFT calculations show that the MH transition is driven by an interplay of correlation effects and a structural (GdFeO$_3$-type)-distortion due to a reduction, not only of the bandwidth, but also of the effective orbital degeneracy \cite{pavarini}. This indicates that in distorted CaVO$_3$, the $t_{2g}$ degeneracy is lifted on the surface, leading to stronger spectral weight changes as observed in our photon energy dependent PE results.

It is also noted that the spectral intensity at $E_F$ can be suppressed by Coulomb interactions. This
was shown by Monte-Carlo calculations \cite{tomita} of spectral changes across the MH transition for a three-dimensional Hubbard model. 
The results indicated a 4-peak structure in the DOS as a function of temperature, with the quasi-particle peak at $E_F$ splitting into two peaks at low temperature, in addition to the presence of upper and lower incoherent peaks \cite{tomita}. The calculated results suggest similarity with the pseudogap-like behavior in the quasi-particle peak observed in our PES results. However, this theoretical calculation assumes an intermediate correlation regime \cite{tomita}, rather than the strongly correlated regime. This is in contrast to the fact that SrVO$_3$ and CaVO$_3$ are usually considered as strongly correlated systems as estimated from PE and inverse PE spectra, with $U \sim$ 4 eV and $W \sim$ 2.5 eV \cite{morikawa95}.

In conclusion, we studied the bulk and surface ES of SrVO$_3$ and CaVO$_3$ with very high resolution using low energy (7 - 21 eV) PES. The coherent spectral weight of SrVO$_3$ and CaVO$_3$ shows a systematic suppression up to 0.2 eV-binding energy on reducing incident photon energy from 21 to 7 eV, due to increasing bulk sensitivity. The bulk ES of SrVO$_3$ and CaVO$_3$ are similar. The results show evidence for the $t_{2g}$ degeneracy being lifted on the surface, and a separation of  $d_{xy}$ and $d_{yz,zx}$ derived states from the bulk $t_{2g}$ feature. The stronger distortion on the surface of CaVO$_{3}$ compared to SrVO$_{3}$ is directly reflected in the coherent DOS at $E_F$, consistent with recent theory. 

\begin{acknowledgments}
We thank Professor K. Nasu for valuable comments. This work was
supported by grants from: the Ministry of Education, Culture Sports, Science and Technology of Japan; grant No. 13CE2002 of MEXT Japan; and the Cooperative Research Program of HiSOR (Accept No. 04-A-7).
\end{acknowledgments}


\end{document}